\documentclass{JHEP3}

\newcommand{\bbibitem}[1]{\bibitem{#1}\marginpar{#1}}

\def\Label#1{\label{#1}%
  \smash{\hbox to0pt{\raise1ex\hbox{\tiny[#1]}\hss}}}
\def\noLabels{\let\Label=\label}
\def\nobbibitem{\let\bbibitem=\bibitem}

\title{
Information recovery from black holes
}

\author{Vijay Balasubramanian \\ David Rittenhouse Laboratories, University of
Pennsylvania, \\Philadelphia, PA 19104, USA \\
\texttt{vijay@physics.upenn.edu}}

\author{Donald Marolf  \\
Physics Department, UCSB, Santa Barbara,  \\ CA 93106, USA \\
\texttt{marolf@physics.ucsb.edu}}

\author{Moshe Rozali \\
Department of Physics and Astronomy, University of British
Columbia, \\ Vancouver, BC
V6T 1Z1, Canada\\\it{and}\\
 Perimeter
Institute, \\ Waterloo, ON N2L 2Y5, Canada\\
\texttt{rozali@physics.ubc.ca, mrozali@perimeterinstitute.ca}}

\abstract{We argue that if black hole entropy arises from a finite
number of underlying quantum states, then any particular such
state can be identified from infinity.    The finite density of
states implies a discrete energy spectrum, and, in general, such
spectra are non-degenerate except as determined by symmetries.
Therefore, knowledge of the precise energy, and of other commuting
conserved charges, determines the quantum state.   In a
gravitating theory, all conserved charges including the energy are
given by boundary terms that can be measured at infinity.   Thus,
within any theory of quantum gravity, no information can be lost
in black holes with a finite number of states.   However,
identifying the state of a black hole from infinity requires
measurements with Planck scale precision.  Hence  observers with
insufficient resolution will experience information loss. }

\date{February 2006}


\preprint{hep-th/0604045\\UPR-T-1150}

\begin{document}


\section{Introduction}
\label{intro}

There are two central questions concerning the quantum physics of
black holes.  First, why do classical black holes apparently have a
finite entropy equal to a quarter of the horizon area?   Second,
does information escape from an evaporating black hole, and if so,
how?    One might suppose that the answers to these two questions
would require separate inputs from new physics.   However, here we
argue that any quantum mechanical theory of gravity that explains
the finite entropy of black holes as the coarse grained description
of a large number of microstates must also permit measurement of
these states at infinity.

The basic argument is simple. If a black hole represents a finite
number of states, $N \sim e^{S_{BH}}$, then the energy spectrum of
the black hole must be discrete. In general such discrete spectra
are quantum-mechanically non-degenerate, except as determined by
symmetries of the system.  Knowledge of the precise energy, along
with the other commuting conserved charges,  thus determines the
quantum state.    But since, in generic gravitating theories,
charges (such as the energy) are given by boundary terms, this leads
to the remarkable conclusion that complete knowledge of the black
hole state is contained in the asymptotic region\footnote{For some
(singular) extremal black holes, this conclusion is also implied by,
but not dependent on, the recent observation that these spacetimes
admit a classical moduli space of non-singular, horizon-free
supergravity microstates which respond to most probes as if they are
singular black holes \cite{Mathur,LLM,Babel,masaki}.}.

Below we elaborate upon this observation, and argue that the
relevant asymptotic measurements will always involve either very
short distances that vanish as $\hbar \to 0$ or very long time
scales that diverge in this limit.    Either way, a conventional
{\it local, classical} observer cannot measure the internal state of
a black hole, although the information is present in the asymptotic
region, and can be measured by observers with more sensitive
instruments. We also explore why ``internal observables'' containing
information inaccessible to the asymptotic observer do not exist,
despite their apparent presence in effective field theory.

\section{Information recovery}
\label{access}

Uncharged black holes in asymptotically flat space do not come to
equilibrium with their radiation and eventually evaporate
completely. Thus, in order to discuss information recovery from
black holes it is helpful to begin by placing black holes ``in a
box'' so that an equilibrium configuration of a black hole
microstate accompanied by a bath of radiation can exist.   The
covariant method of achieving this is to consider black holes in a
universe with a negative cosmological constant.

An infrared cutoff, such as the one produced by the curvature
arising from a negative cosmological constant, removes the obvious
continuum in the spectrum of fields associated with translational
symmetry.   Thus, also assuming that black holes have a finite
number of microstates, the entire gravitating system has a finite
number of states below any energy $E$.   Due to the rapid growth of
the Bekenstein-Hawking entropy, one expects that at sufficiently
high energies the entropy of such systems is dominated by black
holes. The typical state then involves a very heavy black hole in
equilibrium with a small amount of radiation.

  We will argue that complete information concerning the microstates of such black
holes is available in the asymptotic region of spacetime. The central point is that,
in a generic gravitating system, the energy is determined at
infinity.    As with the familiar ADM energy in asymptotically flat
spacetimes, bulk contributions to the energy vanish due to the
gravitational constraints.  Thus, the energy is given entirely by
a surface term.

We begin by discussing black holes in energy eigenstates.  In
quantum mechanics, a discrete spectrum is generically
non-degenerate, except as determined by symmetries.  Thus, a precise
measurement of the energy (and other conserved charges) of a black
hole spacetime is sufficient to identify any particular energy
eigenstate.

 However, such a measurement will require precision that grows exponentially in $1/\hbar$.
 To see this, we temporarily ignore any additional conserved charges. Let us now suppose
 that a measuring device with an  energy resolution $\Delta E$ interacts with a gravitating
 system of total energy $E$ and  entropy $S(E)$. By the usual statistical mechanical
 understanding of entropy, this means that the number of states between $E$
 and $E + \Delta E$ is $\sim e^S$.  Since our system has a non-degenerate spectrum, the
 energy level spacing between $E$ and $E + \Delta E$ must be
\begin{equation}
\delta E \sim \Delta E \, e^{-S} \, .
\end{equation}
In the black hole dominated regime,  the density of states is given by
\begin{equation}
\frac{dN}{dE} \approx \frac{d e^{S_{BH}}}{dE} = e^{S_{BH}} \frac{dS_{BH}}{dE} \, ,
\end{equation}
where $S_{BH}$, the black hole entropy, grows as some power of the
energy $E$. A measuring device with an energy resolution $\Delta E$
will interact with
\begin{equation}
N(E) \, \Delta E \approx e^{S_{BH}} \, \times \Delta E \, {dS_{BH} \over dE}
\end{equation}
states.   The associated entropy,
\begin{equation}
\ln(N(E) \, \Delta E) = S_{BH} + \ln(\Delta E) + \ln\left( {dS_{BH} \over dE } \right) \, ,
\end{equation}
is same as the entropy of the black hole up to logarithmic
corrections. Thus the coarse-grained entropy measured by a device
with resolution $\Delta E$ will equal the black hole entropy to
leading order.

Choosing $\Delta E$ to be any power law in $E$ provides a coarse-graining which gives rise to
the large entropy associated with the black hole.  Nevertheless,
measuring the energy of the state with a much greater precision $\delta E \sim \exp(-S_{BH})$
would determine either a single state, or a small number of degenerate states which can be
identified through measurements of other conserved charges.  However, the Heisenberg
uncertainty principle dictates that any such measurement must extend over an enormous
length of time:
\begin{equation}
\label{time}
\delta t \sim \frac{1}{\delta E} \sim \exp(S_{BH}).
\end{equation}
In the classical limit, $\hbar \to 0$ or $\ell_p \to 0$, this timescale diverges
exponentially because
\begin{equation}
S_{BH} = {A \over 4 \, \ell_p^2} +  \, {\rm corrections} \, .
\end{equation}
The timescale (\ref{time}) is comparable to the system's Heisenberg
recurrence time, over which a generic state in the interval $\Delta
E$ develops a matrix element of order one with any other such
state\footnote{The importance of recurrence times in discussions of
gravitational entropy has been highlighted in, e.g.,
\cite{EBHAdS,recur1,recur2,recur3,recur4,recur3b}.  The Heisenberg
time is discussed in \cite{recur3b}.}. It is also the timescale over
which large thermal fluctuations may occur, perhaps replacing the
black hole by a ball of expanding hot gas. While the the gas will
re-collapse to form another black hole on a (relatively) short
timescale, in the meantime it is plausible that the details of the
black hole's internal state are clearly visible from infinity. Thus,
in retrospect it is perhaps not surprising that an experiment
lasting a time $\delta t  \sim \exp(S_{BH})$ can identify the
internal state of a black hole\footnote{It has been shown that the
states of certain (singular) black holes can be detected by
asymptotic measurements made over such exponentially long timescales
\cite{masaki}.}.

We can now consider the possibility of additional conserved charges
that commute with the energy. Charges associated with gauge
symmetries (e.g., angular momentum, electromagnetic charges etc.)
can clearly be measured at infinity in the usual ways.  While any
charges that are not coupled to long range gauge fields could result
in degeneracies that cannot be disentangled, such degeneracies will
be small because representations of typical symmetry groups do not
grow exponentially quickly\footnote{In fact,  such charges can
sometimes be measured from infinity. Examples include the asymptotic
detection of black hole ``hair" arising from underlying discrete
symmetries (e.g. \cite{discretecharge}), supersymmetry \cite{masaki}
and integrability \cite{integrability}.}.    The dominance of the
energy can also be seen from the fact that, in black hole
thermodynamics, fixing both the mass $M$ and taking the angular
momentum to vanish leads to the same entropy (to leading order) as
specifying only the mass $M$ and leaving the angular momentum
unconstrained.   Thus the overwhelming majority of the information
is available at infinity in the energy spectrum.

So far we have discussed black holes in energy eigenstates.  We now turn to general
superpositions
\begin{equation}
|\psi \rangle = \sum_n a_n |E_n \rangle \, .
\end{equation}
As is always the case for quantum systems, one cannot experimentally
determine the values of all the coefficients $a_n$ given only a
single black hole on which to perform measurements.   Thus, to
demonstrate whether the information is available at infinity, we
should ask whether one can measure the $a_n$ to arbitrary accuracy
given a large number of black holes prepared identically in the
state $| \psi \rangle$. The frequency with which measurement of the
energy gives the result $E_n$ yields the magnitudes $|a_n|^2$. What
remains is to obtain phase information.   As usual, phase
information can be associated to measurement of an operator ${\cal
B}$ which does not commute with the energy, i.e., time-dependent
operators.  Examples of such observables are boundary values of
fields at asymptotic infinity.

For simplicity, let us choose the state $|\psi \rangle$ to be a
superposition of only two energy eigenstates $|\psi \rangle = a_1
|E_1 \rangle + a_2 |E_2 \rangle$.    If we now repeatedly measure
the value of ${\cal B}$ we obtain its eigenvalues $B_n$ with some
frequency.  Consider two such eigenvalues $B_1$ and $B_2$ and the
associated eigenstates
\begin{eqnarray}
|B_1 \rangle &=& \sum_n b_{1,n} |E_n \rangle \, , \nonumber \\
|B_2 \rangle &=& \sum_n b_{2,n}|E_n \rangle \, .
\end{eqnarray}
The coefficients $b_{m,n}$ are determined by the underlying theory; we take to be known
quantities.
The relative frequencies of measurement of $B_1$ and $B_2$ are determined by the overlaps
\begin{eqnarray}
|  \langle B_1 | \psi \rangle |^2 &=& |a_1 \, b_{1, 1} +  a_2 \, b_{1, 2}|^2 \nonumber  \\
| \langle B_2 | \psi \rangle |^2 &=&  |a_1 \, b_{2, 1} +  a_2 \, b_{2, 2}|^2.
\label{decompose}
\end{eqnarray}
The ratio of these two frequencies depends on both the magnitudes
and the phases of $a_1$ and $a_2$.   The phase dependence arises
because $[{\cal B},H] \neq 0$ and therefore the $b_{i,j}$ are
generically non-zero.  Since the $b_{i,j}$ are known, and the
magnitudes $|a_i|$ were already determined by measurements of the
energy, the relative frequency of $B_1$ and $B_2$ allows us to
ascertain the relative phase of $a_1$ and $a_2$.      For a more
general superposition $|\psi \rangle$, repeating similar
measurements fully determines the ray in Hilbert space.    In this
sense, full information about the microstate is available outside
the black hole\footnote{  The picture of information recovery
offered here {\it differs} significantly from the idea explored in
\cite{EBHAdS,hawking,recur3,recur4,recur3b} that  a summation over
multiple classical saddle points with the same asymptotic boundary
conditions might allow for information recovery.   Indeed,
\cite{recur3,recur4,recur3b} showed in explicit examples that this
mechanism was insufficient.  Rather, our perspective is consistent
with \cite{Babel}, where information is lost simply by the erasure
of quantum mechanical detail in semiclassical measurements.}.

The chief difficulty in extending the above reasoning to
asymptotically flat spacetime is that the translational symmetry
results in continuous spectra.   We will nevertheless assume that it
is possible to interpret the entropy of a black hole in
asymptotically flat space in terms of a finite number of
microstates, perhaps by explicitly considering a black hole in a
box, or by otherwise restricting attention to the local region
surrounding the black hole.    In this region we can consider both
the black hole and the thermal atmosphere that it generates as it
evaporates.   Within this framework, there should be an approximate
notion of energy and we should again expect it to have a discrete,
non-degenerate spectrum.    Given this discreteness, measurements
analogous to those described above will identify the black hole
microstates.

\section{The absence of an unobservable interior}
\label{observables}

We have argued that, in any quantum mechanical theory of gravity in
which black holes have a finite number of internal states, one
expects all information about the state to be available near
infinity. This conclusion can be stated in terms of the commutator
of operators:  because the spectrum of the Hamiltonian is
non-degenerate, {\it all} observables which commute with the
Hamiltonian are in fact functions of the Hamiltonian itself.

 There may appear to be a tension between this observation and the fact that, in
 classical general relativity, there are independent observables localized inside
 the black hole.  Because the interior is causally separated from infinity,  such observables
 commute with all asymptotic quantities. However, in our picture this is an artifact of
 the strict classical limit.   Recall that the classical description of black holes in
 the $\hbar \to 0$ limit replaces very long time-scales of order $e^{1/\hbar}$ by infinity.
 Thus probe measurements of the sort necessary to resolve the states of a black hole are
 unavailable in the classical limit.  As such, the usual picture of the black hole
 with a causally disconnected interior is the correct effective classical description. Even
 when $\hbar \neq 0$  this remains the effective description for semiclassical observers
 lacking the measurement precision necessary to resolve the microstates.

Similarly, one can readily imagine that for some class of operators
$\{ \cal O \}$, the commutators with Hamiltonian $H$ simply vanish
rapidly in the classical limit, leading to an approximate notion of
a causally separated region.    However,  in our picture none of
these commutators vanishes exactly for $\hbar \neq 0$.   How might
this be explained in semiclassical terms?  Consider local quantum
field theory on a fixed black hole spacetime. In this context, there
are observables $L$ that are localized inside the black hole. To
promote these operators to observables in quantum gravity one must
make them diffeomorphism invariant.   Procedures to achieve this,
such as integrating $L$ suitably over spacetime, generally lead to
non-local operators which, when evaluated on particular spacetimes,
receive contributions only from a small region \cite{GMH}.   In the
classical limit this region will be contained inside the horizon.
However, at finite $\hbar$ there is always some spread, which
plausibly leads to non-vanishing commutators with operators near
infinity.   This may be related to rare large fluctuations of the
black hole to a thermal-gas like state which are expected over the
recurrence time  (\ref{time}) and which cause the horizon to be
ill-defined when $\hbar \neq 0$.

\section{Discussion}
\label{disc}

 We have argued that if black hole entropy arises from a finite number of underlying
 quantum states, then, in any quantum mechanical theory of gravity, the information
 needed to identify a particular microstate is available at infinity.    We used the
 fact that, in a generic gravitating theory, the energy is given by a surface term at
 infinity.  While new physics is needed to explain why a given black holes is associated
 with a finite number of states, no further new physics is required to make information
 about black hole states available at infinity.

Although the information is available in our sense, there may be
practical or even in-principle limitations to recovery of the
information by a physical apparatus.  For example, in asymptotically
flat space, one must also deal with the fact that black holes
represent broad resonances as opposed to sharp energy eigenstates
\cite{MS}.  Even for stable black holes it is clear that, in order
to separate black hole microstates, a measurement apparatus will
itself require a large number of internal states\footnote{Issues
associated with such measuring devices in gravity have been
discussed in \cite{recur2,GMH}.}.    In order to minimize
back-reaction of such a system, one would need to either dilute it
in space, or move it far away.  Either way, the interactions of the
apparatus with the black hole would be weakened, making the
practical task of state identification more difficult.

\subsection*{Acknowledgments}
We have enjoyed useful conversations with David Berenstein,
Bartlomiej Czech, Jan de Boer, Veronika Hubeny, Vishnu Jejjala,
Matt Kleban, Klaus Larjo, Rob Myers, Massimo Porrati, Mukund
Rangamani, Simon Ross, Joan Simon, and Mark Srednicki.  V.B. is
supported in part by the DOE under grant DE-FG02-95ER40893, by the
NSF under grant PHY-0331728 and by an NSF Focused Research Grant
DMS0139799. D.M. was supported in part by NSF grants PHY0354978
and PHY99-07949, by funds from the University of California, and
by funds from the Perimeter Institute of Theoretical Physics. M.R.
is supported by a discovery grant from NSERC of Canada, and by
funds from the Perimeter Institute. Research at the Perimeter
Institute is supported in part by funds from NSERC of Canada and
MDET of Ontario.

This paper was begun at the Perimeter Institute ``Summer School on
Strings, Gravity and Cosmology'' and completed at the Arnold
Sommerfeld Center's  workshop on ``Black Holes, Black Rings and
Topological Strings''.   V.B. and D.M. are grateful to the
organizers of both workshops for their hospitality.  D.M. would
also like to thank the Kavli Institute for Theoretical Physics for
its hospitality during intermediate stages of this work.

\end{document}